\documentstyle[jkas2]{article}

\runningauthor{KANG, LAKE, AND RYU}
\runningtitle{GLOBULAR CLUSTER FORMATION}
\beginpage{1}
\endpage{11}

\def\etal{{\it et al.~}}
\def\eg{{\it e.g.,~}}
\def\ie{{\it i.e.~}}
\def\Htwo{${\rm H_2}~$}
\def\H{{\rm H}}
\def\He{{\rm He}}
\def\K{~{\rm K}}
\def\g{~{\rm g}}
\def\kms{~{\rm km~s^{-1}}}
\def\cm3{~{\rm cm^{-3}}}
\def\yrs{~{\rm yrs}}
\def\pc{~{\rm pc}}
\def\kpc{~{\rm kpc}}
\def\Msun{~{\rm M}_{\sun}}

\def\lsim{\mathrel{  %% \mathrel makes \lsim look like other relation symbols
        \raise0.3ex\hbox{$<$}\kern-0.75em{\lower0.65ex\hbox{$\sim$}}}}
\def\gsim{\mathrel{
        \raise0.3ex\hbox{$>$}\kern-0.75em{\lower0.65ex\hbox{$\sim$}}}}

\begin{document}

\title{FORMATION OF PROTO-GLOBULAR CLUSTER CLOUDS BY THERMAL INSTABILITY}

\author{HYESUNG KANG$^1$, GEORGE LAKE$^2$, AND DONGSU RYU$^3$}

\address{$^1$Department of Earth Sciences, Pusan National University, Pusan
            609-735, Korea \\
         $^2$Department of Astronomy, University of Washington, Box 351580,
             Seattle, WA 98185-1580, USA\\
         $^3$Department of Astronomy \& Space Science, Chungnam National 
             University, Daejeon 305-764, Korea\\
{\it E-mail: kang@uju.es.pusan.ac.kr,  lake@heremes.astro.washington.edu,
             and ryu@canopus.chungnam.ac.kr }}

\address{\normalsize{\it (Received July ??, 2000; Accepted ???. ??,2000)}}

\abstract{
Many models of globular cluster formation assume the presence of
cold dense clouds in early universe.
Here we re-examine the Fall \& Rees (1985) model for formation of
proto-globular cluster clouds (PGCCs) via thermal instabilities
in a protogalactic halo.
We first argue,
based on the previous study of two-dimensional numerical simulations of 
thermally unstable clouds in a stratified halo of galaxy clusters 
by Real \etal (1991), 
that under the protogalactic environments
only nonlinear ($\delta \ga 1$) density inhomogeneities can condense into
PGCCs without being disrupted by the buoyancy-driven dynamical instabilities.
We then carry out numerical simulations of the collapse of overdense clouds in
one-dimensional spherical geometry, including self-gravity and radiative
cooling down to $T=10^4 \K$.
Since imprinting of Jeans mass at $10^4\K$ is essential to this model,
here we focus on the cases where external UV background radiation prevents
the formation of \Htwo molecules and so prevent the cloud from cooling below
$10^4\K$. 
The quantitative results from these simulations can be summarized as follows: 
1) Perturbations smaller than $M_{\rm min} \sim (10^{5.6}\Msun) (n_h/0.05\cm3)^{-2}$
cool {\it isobarically}, where $n_h$ is the unperturbed halo density, while
perturbations larger than $M_{\rm max}\sim (10^8\Msun) (n_h/0.05
\cm3)^{-2}$ cool {\it isochorically} and thermal instabilities do not operate.
On the other hand, intermediate size perturbations 
($M_{\rm min}<M_{\rm pgcc}<M_{\rm max}$)
are compressed {\it supersonically}, accompanied by strong accretion shocks.
2) For supersonically collapsing clouds,
the density compression factor after they cool to $T_c=10^4\K$ range
$10^{2.5}-10^6$, while the isobaric compression factor is only $10^{2.5}$.
3) Isobarically collapsed clouds ($M<M_{\rm min}$) are too small to be
gravitationally bound. 
For supersonically collapsing clouds, however,
the Jeans mass can be reduced to as small as
$10^{5.5} \Msun (n_h/0.05 \cm3)^{-1/2}$ at the maximum compression
owing to the increased density compression.
4) The density profile of simulated PGCCs can
be approximated by a constant core with a halo of
$\rho \propto r^{-2}$ rather than a singular isothermal sphere.}


\maketitle

\section{INTRODUCTION}

Globular clusters (GCs) are the oldest relics in our galaxy, presenting 
a cosmological challenge, as their ages often burst the allowed bounds
of the expansion age. 
Since there are {\it at least} three
distinct populations of GCs, the old halo and 
the middle-aged disk populations of GCs in our Galaxy (Zinn 1985), 
and the young GCs in interacting galaxies (Ashman \& Zepf 1992),
we may well need more than one model to explain their formation and evolution. 
In the present study, we focus on the formation of the halo GCs.
Half of the total mass of our galaxy appears to be contained within $\sim 50\kpc$, 
whereas roughly half of the luminosity of the disk as well
as half of the total number of halo GCs are distributed within
the solar galactocentric distance ($\sim 8\kpc$).
So halo GCs clearly know about the {\it dissipation} that followed galaxy
formation and we must place their formation in the protogalactic context.
This encourages models that form halo clusters during the formation
of the protogalaxy, such as thermal instabilities in the
protogalactic halo (Fall \& Rees 1985) or pregalactic
cloud collisions (Gunn 1980; Lake 1987; Kang \etal 1990;
Kumai, Basu, \& Fujimoto 1993).

Many models of globular cluster formation assume the presence of 
{\it proto-globular cluster clouds} (PGCCs) in pressure equilibrium 
with hot halo gas (\eg  Gunn 1980; Dopita \& Smith 1986; 
Brown, Burkert \& Truran 1991; Kumai, Batsu, \& Fujimoto 1993).
The model by Fall \& Rees (1985, FR85 hereafter) has been most widely adopted 
to explain the formation of such PGCCs:
the thermal instability drives catastrophic cooling of over-dense regions 
in the hot halos of protogalaxies and condenses them to dense clouds.
In the present study, adopting FR85 model, we follow the thermal 
and dynamical evolution of over-dense clouds by numerical simulations 
in order to examine more
quantitatively the physical properties of resulting PGCCs.

Although the thermal instability was extensively investigated in several
areas of astrophysics, in recent years it has received much attention
especially for X-ray cluster cooling flows: 
({\it cf.} Balbus 1986; Malagoli, Rosner \& Bodo 1987; 
Balbus \& Soker 1989; Malagoli, Rosner, \& Fryxell 1990). 
According to the results from numerical simulations in 
one-dimensional (1D) 
geometries (David, Bregman \& Seab 1988 [DBS88 hereafter]; Brinkmann, 
Massaglia \& M\"uller 1990 [BMM90]),
the over-density undergoes a {\it quasi-static compression} 
in near pressure equilibrium  
when a cloud is small enough to adjust to pressure change faster than 
it cools, while larger clouds may undergo a ``supersonic compression stage'',
leading to a density increase much higher than what is expected
from the isobaric compression.
The plane-parallel calculations by DBS88 showed that the
density contrast increases to the ``isobaric" ratio of the temperature
of the hot background medium ($T_h\sim 10^7-10^8\K$) and
the minimum temperature ($T_c\sim 10^4\K$) where the cooling becomes
inefficient.
On the other hand, the spherically symmetric calculations of BMM90 showed
that the density contrasts can be more than three orders of magnitude higher
than those found in the plane-parallel calculations of DBS88.
The difference between two geometries is mainly due to the fact that the
gravitational acceleration increases as $GM/R^2$ when the spherical
cloud collapses,
while it remains the same (independent of $R$) when the plane sheet
compresses.
Also isotropic spherical compression (focusing) leads to much higher
density accelerating the cooling and rise in density.
However, 2D simulations of the collapse of an
elongated blob by BMM90 showed a shape instability
to a pancake-shape leading to evolution best represented by
the 1D plane-parallel case.
Thus geometry is clearly important in the nonlinear evolution.
On the other hand, 2D simulations
by V\'azquez-Semadeni, Gazol, \& Scalo (2000)
showed that the condensation of fluctuating density field
via thermal instability leads to a network of transient filaments
which later accrete onto isolated round blobs.
These blobs form at intersections between filaments just like
galaxy clusters in large-scale structure formation in cosmological
simulations.
This implies that the 1D spherical simulation can actually
represent a generic collapse of the highest density peaks
in the randomly fluctuating density field in 3D.

Numerical simulations in a 2D stratified background 
(Yoshida, Hattori, \& Habe 1991; Reale \etal 1991 and
references therein), on the other hand, have shown that
the dynamic instabilities, such as Rayleigh-Taylor and Kevin-Helmholtz 
instabilities due to buoyancy-driven oscillations 
can disrupt the thermally unstable bubbles
if the timescale for buoyancy oscillation (Brunt-V\"as\"al\"a period) 
is shorter than the local radiative cooling time. 
Thus it is believed that the classic theory of thermal instability can not 
explain the effective mass depletion from the cooling
flows of galactic clusters.
We will examine the role of such dynamic instabilities
for the formation of PGCCs in the protogalactic halo environment 
in the next section. 
The details of our models are given also in
this section.  The main results are in \S 3.
Conclusions are given in \S 4.

\section{MODEL CALCULATIONS}

\begin{figure}
\vspace{-0.2truein}
\epsfysize=3.4in\epsfbox[105 210 480 580]{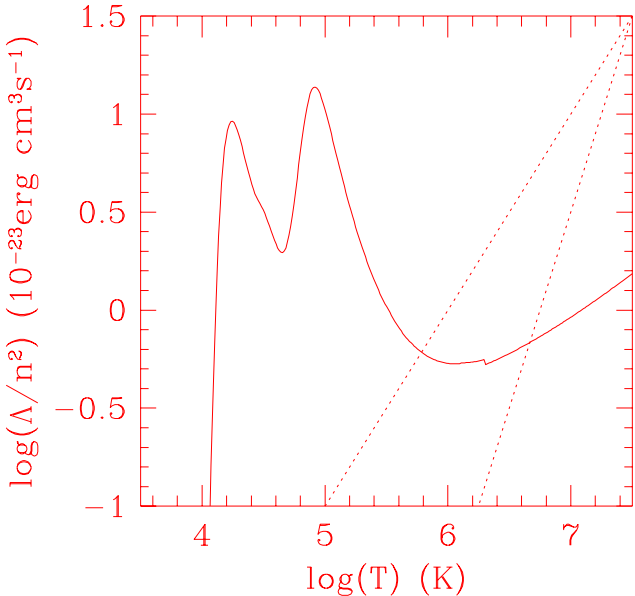}
\vspace{-0.2truein}
\caption{The equilibrium cooling rate ($f(T) = \Lambda/n^2$) for a primordial
gas composed of H and He is given in units of $10^{-23} {\rm erg~cm^3~s^{-1}}$.
Dotted lines show $d\ln(\Lambda/n^2)/d\ln T=1$ and
$d\ln(\Lambda/n^2)/d\ln T=2$.}
\vspace{0truein}
\end{figure}

\subsection{Protogalactic Halo Model}

We adopt an idealized model for the protogalactic halo gas as in FR85;
an isothermal sphere based on the Milky Way's properties, with a
circular velocity $V_c=220 \kms $
and corresponding gas temperature $T_h = 1.7\times 10^6\K$. 

Many previous studies used a 
gas distribution specified by the condition that the 
free fall time to the center of the protogalaxy 
equals the cooling time, yielding the
halo gas density at a distance $R_g$ from the center
$\rho_1 \sim 1.7\times 10^{-25}\g \cm3 (R_g / 10\kpc)^{-1}$. 
The use of $\rho_1$ has lead previous workers to consider 
halo densities of $n_h=0.1 \cm3$ (FR85; Kang \etal 1990) 
or even $n_h=1 \cm3$ (Vietri \& Pesce 1995). For a primordial
gas of $\H$ and $\He$ with an assumed
ratio $n(\He)/n(\H)=0.1$, the halo gas mass density is 
$\rho_h=(2.34\times 10^{-24}\g \cm3)n_h$. For the isothermal
sphere in hydrostatic equilibrium, however, the 
gas density is given by 
$\rho_2= 5.8\times 10^{-25} \g \cm3 f_B (R_g /10\kpc)^{-2}$,
where $f_B$ is the ratio of the baryonic to total matter density.
Realistic values should be $0.02\la n_h \la 0.07 \cm3$ for
$ R_g \sim 10\kpc$ (assuming $f_B\sim 0.1$)
with some degree of central concentration. A value as high
as $n_h=1 \cm3$ is greater than the total density of baryonic and dark
matter.
Thus we take $0.05 \cm3$ as a fiducial value and 
express the halo gas density as 
\begin{equation}
n_h \approx (0.05 \cm3) R_{10}^{-q},
\end{equation}
where $R_{10}$ is $R_g$ in units of $10\kpc$ and
$1\la q \la 2$.

Although the density irregularities in protogalaxies likely result
from a superposed spectrum of perturbations, 
we consider, for simplicity, isolated spherically symmetric clouds 
resulted from sinusoidal perturbations. 
So the initial gas density of the cloud $n_{cloud}$
decreases gradually with radius $r$ from its center
to its edge $R_c$ as follows:
\begin{equation}
n_{cloud} = n_h \left[1.0+ \delta \cos\left({\pi\over2}{r \over R_c}\right)
\right]~~{\rm for}~r<R_c~~
\end{equation}
where $\delta$ is the amplitude of the initial
density perturbation and $R_c$ is the cloud radius. 

The initial temperature throughout the cloud is set by the condition of
hydrostatic equilibrium.
To start in the nearly linear regime, we always begin with 
$\delta\sim1$, unless otherwise noted.
Initial parameters of the model clouds are summarized in Table 1.

%------------------------------------------------ TABLE 1 START
\begin{table*}
\begin{center}
{\bf Table 1.}~~Initial Parameters for Model Clouds\\
\vskip 0.3cm
\begin{tabular}{ lrrrrrr }
\hline\hline
Model & $n_h$ & ~~~~~$\delta$ & ~$R_c/l_{cool}$ $^{\rm a}$ & ~~~~~~~~$R_c$ &
~~~~~$M_c$   \\
~ & $(\cm3)$ & ~ & ~ & (pc) & (M$_{\sun}$) & flow motion$^{\rm b}$ \\

\hline
SM1  & 0.1 &1 & 0.042& 4.20$\times 10^1$ & 1.4$\times 10^3$ & isobaric\\
SM2  & 0.1 &1 & 0.092& 9.24$\times 10^1$ & 1.5$\times 10^4$ & isobaric\\
SM3  & 0.1 &1 & 0.21& 2.10$\times 10^2$ & 1.8$\times 10^5$ & isobaric\\
SM4  & 0.1 &1 & 0.42& 4.20$\times 10^2$ & 1.4$\times 10^6$ & supersonic \\
SM5  & 0.1 &1 & 0.63& 6.30$\times 10^2$ & 4.9$\times 10^6$ & supersonic \\
SM6  & 0.1 &1 & 1.1& 1.05$\times 10^3$ & 2.3$\times 10^7$ & supersonic  \\
SM7  & 0.1 &1 & 2.1& 2.10$\times 10^3$ & 1.8$\times 10^8$ & isochoric \\
SM8  & 0.1 &5 & 2.1& 2.10$\times 10^3$ & 3.2$\times 10^8$ & isochoric \\
SM9  & 0.1&0.5& 0.42& 4.20$\times 10^2$ & 1.3$\times 10^6$ & isochoric \\

\hline
SM11 & 1.0 &1 &0.17 & 1.71$\times 10^1$ & 9.8$\times 10^2$ & isobaric  \\
SM12 & 1.0 &1 &0.42 & 4.22$\times 10^1$ & 1.4$\times 10^4$ & supersonic  \\
SM13 & 1.0 &1 &0.68 & 6.83$\times 10^1$ & 6.2$\times 10^4$ & supersonic  \\
SM14 & 1.0 &1 &1.4 & 1.37$\times 10^2$ & 4.9$\times 10^5$ & supersonic  \\
SM15 & 1.0 &1 &2.7 & 2.73$\times 10^2$ & 4.0$\times 10^6$ & isochoric  \\
SM16 & 1.0 &1 &5.5 & 5.46$\times 10^2$ & 3.2$\times 10^7$ & isochoric  \\

\hline
SM21 & 0.01 &1 &0.10& 1.02$\times 10^3$ & 2.1$\times 10^6$ & isobaric \\
SM22 & 0.01 &1 &0.12& 1.23$\times 10^3$ & 8.6$\times 10^6$ & isobaric \\
SM23 & 0.01 &1 &0.21& 2.05$\times 10^3$ & 1.7$\times 10^7$ & supersonic \\
SM24 & 0.01 &1 &0.33& 3.28$\times 10^3$ & 6.9$\times 10^7$ & supersonic \\
SM25 & 0.01 &1 &0.41& 4.10$\times 10^3$ & 1.3$\times 10^8$ & supersonic  \\
SM26 & 0.01 &1 &0.61& 6.14$\times 10^3$ & 4.5$\times 10^8$ & supersonic  \\
SM27 & 0.01 &1 &0.82& 8.19$\times 10^3$ & 1.1$\times 10^9$ & isochoric  \\

\hline
\end{tabular}
\end{center}
\end{table*}
%-------------------------------------------------- TABLE 1 END

\subsection{Time Scales}

The free fall time $t_{ff}^{pg}$ is the time it takes a particle 
dropped from $R_g$ to reach the center of the protogalaxy,
\begin{equation}
t_{ff}^{pg}= \left({\pi\over2}\right)\left({R_g \over V_c}\right)
= 5.6 \times 10^7\yrs \left({R_g\over 10\kpc}\right)_.
\end{equation}
The free fall time of a uniform gas cloud including only its 
self-gravity is
\begin{equation}
t_{ff}^{sg}= \left({3\pi \over 32 G \rho}\right)^{1/2} = 
4.4 \times 10^6\yrs 
\left({n\over 100\cm3}\right)^{-1/2}_,
\end{equation}
where $n$ is the cloud density and any dark matter contribution 
is ignored.  

The cooling time for the halo gas of $T_h$ is given by
\begin{equation}
t_{cl,h} = {1.5 n_h kT_h \over \Lambda}
 \approx (4.0\times 10^7\yrs) R_{10}^q,
\end{equation}
where the cooling rate of $\Lambda/n_h^2 = 5.5\times 10^{-24} 
{\rm erg~ cm^3 s^{-1}}$
at $T_h=1.7\times 10^6$ and the halo density given in Eq. (1) are used to
calculate the latter relation. 
For an {\it isobaric} perturbation of the density contrast $\delta$, the
cooling timescale of the perturbed gas becomes
$t_{cl,c} \approx t_{cl,h}(1+ \delta)^{-2}$, 
considering that the cooling
rate is almost constant for temperatures 
within factor of two of $T\sim 10^6\K$.
We define the {\it cooling distance} as the distance over which the
sound wave of the hot gas can travel within the cooling timescale of
the perturbed gas,
\begin{equation}
l_{cool} \approx c_h t_{cl,c} \approx (8.1\times 10^3\pc) R_{10}^q
(1+\delta)^{-2}_,
\end{equation}
where $c_h=198 \kms$ is the sound speed of the hot gas of $T_h$.

The ratio of the cooling time to the self-gravity time for initial
clouds is 
\begin{equation}
\eta = 6.2\times 10^{-2} R_{10}^{q/2}  
\left({1+\delta \over 2}\right)^{-3/2}_.
\end{equation}
For the range of density considered here, $0.01<n_h<1\cm3$,
cooling takes place too fast for self-gravity to be
important initially. 
However, the cooling becomes inefficient once the gas cools to $10^4$K,
and the gravitational time scale decreases as the density increases.
The density in the condensed clouds increases by a factor of 
$ \chi_{isob}=(T_h/T_c)(\mu_c/\mu_h)\sim 350$ for an isobaric case,
where $\mu_c$ and $\mu_h$ are the molecular weights of the cold and
hot gas, respectively.
The density enhancement is even higher for supersonic collapses,
resulting in the cloud pressure higher than the background pressure.
The typical additional enhancement in supersonic collapses is
$10^2-10^{3.5}$ (see the next section).
Thus, self-gravity becomes dynamically
important in late stages.
Also note that self-gravity would be more important for lower
density environments, since  $\eta \propto n_h^{-1/2}$.

The growth of thermal instabilities in a stratified medium is affected
by the buoyancy-driven dynamical instabilities (see below),
so the local buoyancy oscillation period is also important. For
our halo model it can be estimated by
the local scale height $H$ divided by the sound speed $c_h$,
\begin{equation}
t_{b} \sim { H \over c_h} \sim (5.0\times 10^7\yrs) R_{10},
\end{equation}
since $H\sim R_g$ for the power-law density distribution assumed
here. For linear initial perturbation (\eg $\delta <<1$),
the cooling time is about the same as
the buoyancy oscillation period ($t_{cl,c}\sim t_b$) and
the cooling distance is about the size of scale height of the halo
($l_{cool}\sim R_g$).
This means the global stability needs to be considered (Balbus
\& Soker 1989) for such perturbations.

\subsection{Effects of Dynamical Instabilities}

When a bubble-shaped perturbation condenses due to radiative cooling,
it falls along the direction of gravity and so
it is subject to the Rayleigh-Taylor and Kevin-Helmholtz instabilities.
Nonlinear evolution of such perturbations in a stratified halo of X-ray
clusters was followed via numerical simulations (Hattori \& Habe 1990;
Yoshida, Hattori, \& Habe 1991; Reale \etal 1991).
The bubble is disrupted by these instabilities on a buoyancy oscillation
period if it does not cool and condense significantly.
Thus the cooling timescale must be shorter than the buoyancy oscillation
period in order to prevent the disruption of the bubble.
That depends on three parameters in order of importance:
the ratio of the local buoyancy oscillation period to the local cooling 
time, the initial density contrast of the perturbation,
and the ratio of the perturbation radius to the local
scale-height (Reale \etal 1991).
We note here the first and second parameters are not completely independent,
since the cooling time of the perturbed gas is dependent on the initial
density contrast.

Following Reale \etal (1991), we define for our protogalactic halo model,
\begin{equation}
\xi = {t_b \over t_{cl,c}} \sim 1.25(1+\delta)^2 R_{10}^{(1-q)},
\end{equation}
using Eq. (6).
According to their 2D numerical simulations,
for $\xi \approx 1$ the perturbations lead to fragmentation
due to the dynamical instabilities, but the fragments can be thermally
unstable. For $\xi \gsim 5$, the perturbation cools and
condenses so rapidly that the dynamical instabilities do not have time
to disrupt the bubble.
These numerical simulations find that thermal instabilities
cannot be an effective mechanism for mass depletion from the X-ray cluster
cooling flows, since typically $\xi \sim 0.01$ there.
According to Eq. (9), on the other hand,
$\xi \sim 5$ for $\delta\sim 1$
in a protogalactic halo, so the galactic halo provides more favorable
conditions for thermal instabilities.

Thus the initial density contrast of the perturbations
is the next important physical parameter.
Here, we will take the generic view that the protogalaxy was formed 
from the hierarchical clustering of smaller
substructures and contains some clumpiness and turbulences.
We suggest only {\it nonlinear} (\ie $\delta \ga 1$) inhomogeneities
in the protogalactic halo can condense due to
thermal instabilities without being disrupted by the buoyancy-driven
dynamical instabilities.

\subsection{Numerical Method}

The gas dynamical equations including self-gravity and cooling are:
\begin{equation}
{d\rho \over dt} + \rho \nabla u =0,
\end{equation}
\begin{equation}
{du \over dt} = -{1\over \rho} \nabla P + g,
\end{equation}
\begin{equation}
{de \over dt} = -{1\over \rho} \nabla (Pu) + gu - {\Lambda\over \rho}_,
\end{equation}
where $e=(\gamma-1)p/\rho + (1/2)u^2$ is the total energy of the gas
per unit mass and $g$ is the gravitational acceleration.
We use the Piecewise Parabolic Method (PPM) 
(Colella \& Woodward 1984) in one-dimensional spherical geometry
and treat gravity and cooling as source terms.
The gravitational acceleration is softened near the center according to
\begin{equation}
g = - { GM(<r) \over [{\rm max}(r,s)]^2}_,
\end{equation}
where $G$ is the gravitational constant, $M(<r)$ is the mass within
the radius $r$, and $s\sim0.04 R_c$ is the softening length.
In  most previous numerical
simulations of the thermal instability (\eg, the references in \S 1), 
self-gravity was ignored because the 
gravitational time scale was longer
than the cooling time scale and/or the emphasis was on the 
effects of the thermal instability alone.
In our models, the gravitational
time scale is longer than the cooling time scale initially, 
but not at later times.
Thus, we must follow the transition of 
the inward flow from being driven
by background pressure at early times to self-gravity in 
the last stages of collapse.

The simulations start at $t=0$ with the clouds at rest in pressure equilibrium
and cease when the {\it background} gas has cooled to $10^4\K$
(see below).
The standard mirror condition for the reflecting boundary at the center
is used, while the flow is assumed to be continuous across the outer
boundary.

\subsection{Radiative Cooling Rate} 

Our radiative cooling rate $f(T)$ applies to a primordial gas in 
ionization equilibrium ({\it cf.} Shapiro \& Kang 1987).
(The cooling rate per unit volume $\Lambda = n^2 f(T)$ where
$f(T)$ is a function of temperature only.)
The success of FR85 model is hinged on {\it imprinting} of the
gravitationally unstable mass scale of PGCCs at $10^4\K$ which
is similar to the characteristic mass scale of GCs. 
This requires that formation of \Htwo molecules should be delayed 
longer than a free fall time of PGCCs.
Kang \etal (1990) suggested that the globular cluster might have
formed only during the phase when the protogalaxies are bright  
in UV background radiation.
Hence we assume a model in which UV background radiation
prevents formation of \Htwo, so that 
the cooling rate is zero below $10^4\K$ 
and the minimum temperature is set at $10^4\K$.

The criterion for the thermal instability is that the 
over-density continues to cool faster than its surroundings,
requiring $d\ln f(T)/d\ln T<1$ to destabilize
isochoric condensations and $d\ln f(T)/d\ln T<2$ to destabilize isobaric
condensations (Balbus 1986).
Fig. 1 shows $f(T)$ and 
includes dotted lines where $d\ln f(T)/d\ln T=1$ and
$d\ln f(T)/d\ln T=2$.

We ignore the poorly understood process of thermal conduction.  
With the classical Spitzer (1979) conductivity,  thermal
conduction may dominate over the cooling.  
If we assume that the heat flux is {\it saturated} when the
particle mean free path is comparable to the length scale of
the temperature variation, 
the  conductivity is  one or two orders of magnitude smaller 
than the Spitzer value
(see Gray \& Kilkenny 1980).   
Brinkmann, Massaglia \& M\"uller (1990) showed that
flow quantities differ $\la$3\% when they added
the reduced thermal conduction to their simulations of 
hot gas in clusters of galaxies.

\section{Simulation Results}

\subsection{Classification by Cloud Size}

We calculated spherical collapses for a wide range of halo
densities ($0.01\cm3 \le n_h \le 1\cm3$) and cloud sizes
($0.04 \la R_c/l_{cool} \la 2.5$).
Initial parameters of the model clouds are summarized in Table 1,
where $M_c$ is the mass of the cloud contained within the initial
perturbation.

Radiative cooling of the cloud leads to three different types of
flow motions depending on its size relative to the cooling distance,
as recognized in previous studies (FR85, DBS88).
According to our simulations, they can be summarized as follows:
1) $(R_c/l_{cool}) \lsim 0.2$, {\it small} isobaric compression regime,
2) $0.2\lsim (R_c/l_{cool}) \lsim 1-1.5$, {\it intermediate} supersonic 
compression regime,
and 3) $(R_c/l_{cool}) \gsim 1-1.5$, {\it large} isochoric cooling regime.
While in {\it small} clouds the collapse remains isobaric and quasi-static, 
in {\it large} clouds 
the central gas cools almost isochorically and the thermal instability fails.
Examples of these limiting cases are plotted in Fig. 2.
Spherical collapses of clouds with $R_c/l_{cool}\sim 0.04$ and
$R_c/l_{cool}\sim 2$ in the background halo with $n_h=0.1\cm3$ and
$T_h=1.7\times 10^6\K$ are shown.
Although the transitions between types are obviously continuous and gradual,
we classified our models into ``isobaric'', ``supersonic'', and ``isochoric"
types according to the characteristics of flow motions in the simulations.
The last column of Table 1 shows the types.

%-------------------------------------------------- FIG 2 START
\begin{figure*}
\centerline{\epsfysize=20cm\epsfbox{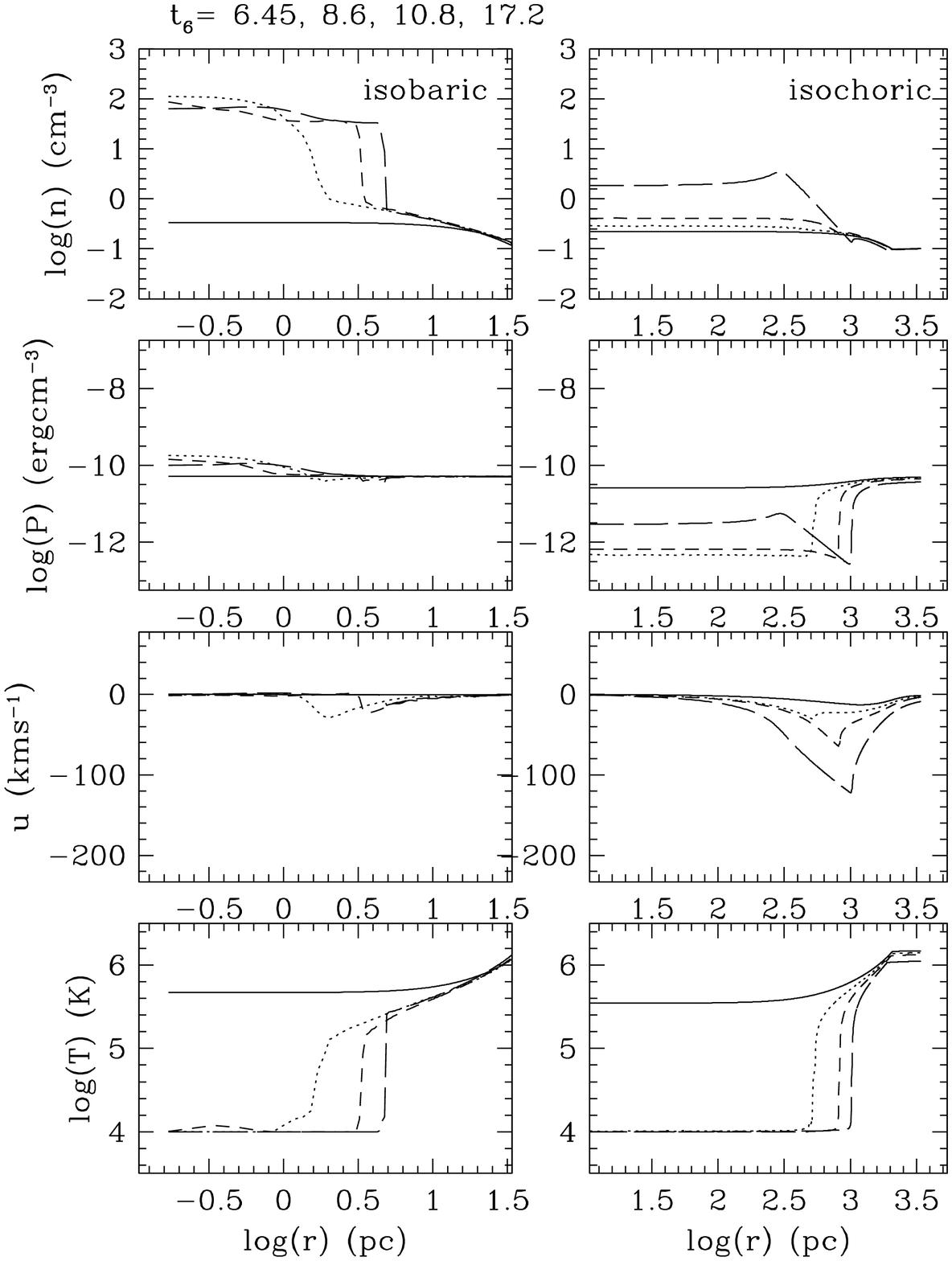}}
\vskip 0.5cm
\hskip 1.7cm {\begin{minipage}{14cm}
{\bf Fig. 2.}---~
Evolution of small {\it isobaric} (model SM1: $R_c=0.04 l_{cool}$, 
left panels) and large {\it isochoric} (model SM7: $R_c=2.1 l_{cool}$, 
right panels)
perturbations in one-dimensional spherical geometry.
Number density, pressure, radial velocity and temperature
are plotted against radius.
The background medium has $n_h=0.1\cm3$ and $T_h=1.7\times 10^6\K$.
The structure is shown at
$t_6=$ 6.45 (solid), 8.6 (dotted), 10.8 (dashed), and 17.2 (long dashed),
where $t_6$ is time in units of $10^6\yrs$.
The small cloud is compressed isobarically with subsonic flow motions (no shock),
while the large cloud cools with only little bit of compression.
\end{minipage}}
\end{figure*}
%------------------------------------------------ FIG 2 END

%-------------------------------------------------- FIG 3 START
\begin{figure*}
\centerline{\epsfysize=20cm\epsfbox{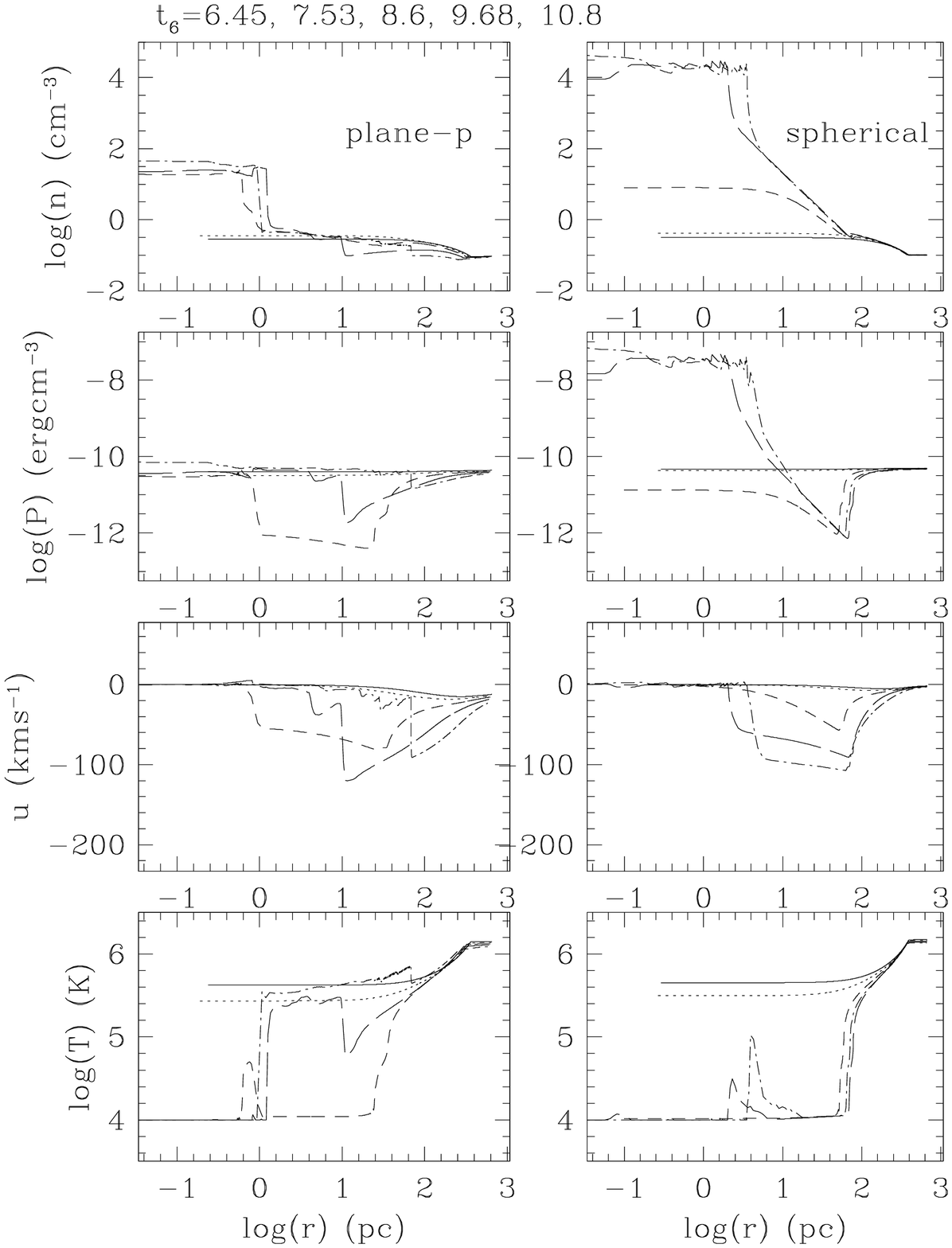}}
\vskip 0.5cm
\hskip 1.7cm {\begin{minipage}{14cm}
{\bf Fig. 3.}---~
Number density, pressure, radial velocity and temperature
are plotted against radius for an {\it intermediate} perturbation of
$R_c=0.4 l_{cool}$
in a background medium of $n_h=0.1\cm3$ and $T_h=1.7\times 10^6\K$.
The panels on the left are for one-dimensional plane-parallel geometry,
while those on the right are for one-dimensional spherical geometry.
The structure is shown at
$t_6=$ 6.45 (solid), 7.53 (dotted), 8.6 (dashed), 9.68 (long dashed) and
10.8 (dot-dashed), where $t_6$ is time in units of $10^6\yrs$.
\end{minipage}}
\end{figure*}
%------------------------------------------------ FIG 3 END
The {\it intermediate} size clouds belong to the most interesting regime.
In order to see the evolution of an intermediate size cloud and the
geometrical effects, spherical collapse as well as plane-parallel collapse
were calculated for one model:
$T_h=1.7\times 10^6\K$, $n_h=0.1\cm3$, $l_{cool}\approx 10^3\pc$,
and the scale height of the plane slab, $H$, and the radius of the
cloud, $R_c$, are $H=R_c=410\pc\sim 0.4 l_{cool}$. 
Fig. 3 shows the evolution at $t/t_{cl,c} = 1.3, 1.5,
1.7, 1.9,$ and 2.2, where $t_{cl,c}=5.0\times 10^6\yrs$.
In both geometries, the gas cools isobarically until $T\sim 2.5\times
10^5 \K$, below which the cooling becomes exponentially rapid and
out of pressure equilibrium.
As the pressure drops in the middle, a compression wave moves in.
The infalling flow accelerates up to $\sim 0.5c_h$,
moves to the center and then bounces back as an {\it accretion shock}.
In the plane-parallel case, the accretion shock
smoothes out the pressure gradient leading to near equilibrium with
the background pressure.
In the spherical case, on the other hand, the pressure of the shocked gas
increases beyond the initial value owing to the central focusing.
Thus the density enhancement factor can be as high as $10^6$ ($n_h \sim
10^4~ \cm3$), which is
$\sim 10^{3.5}$ higher than the isobaric compression ratio.
Although there is a large positive pressure gradient
between the cloud core and the background gas, there exists a halo of
supersonic accretion flow instead of the out-flow,
because the high density and high pressure core 
is gravitationally bound. 

%-------------------------------------------------- FIG 4 START
\begin{figure*}
\centerline{\epsfysize=20cm\epsfbox{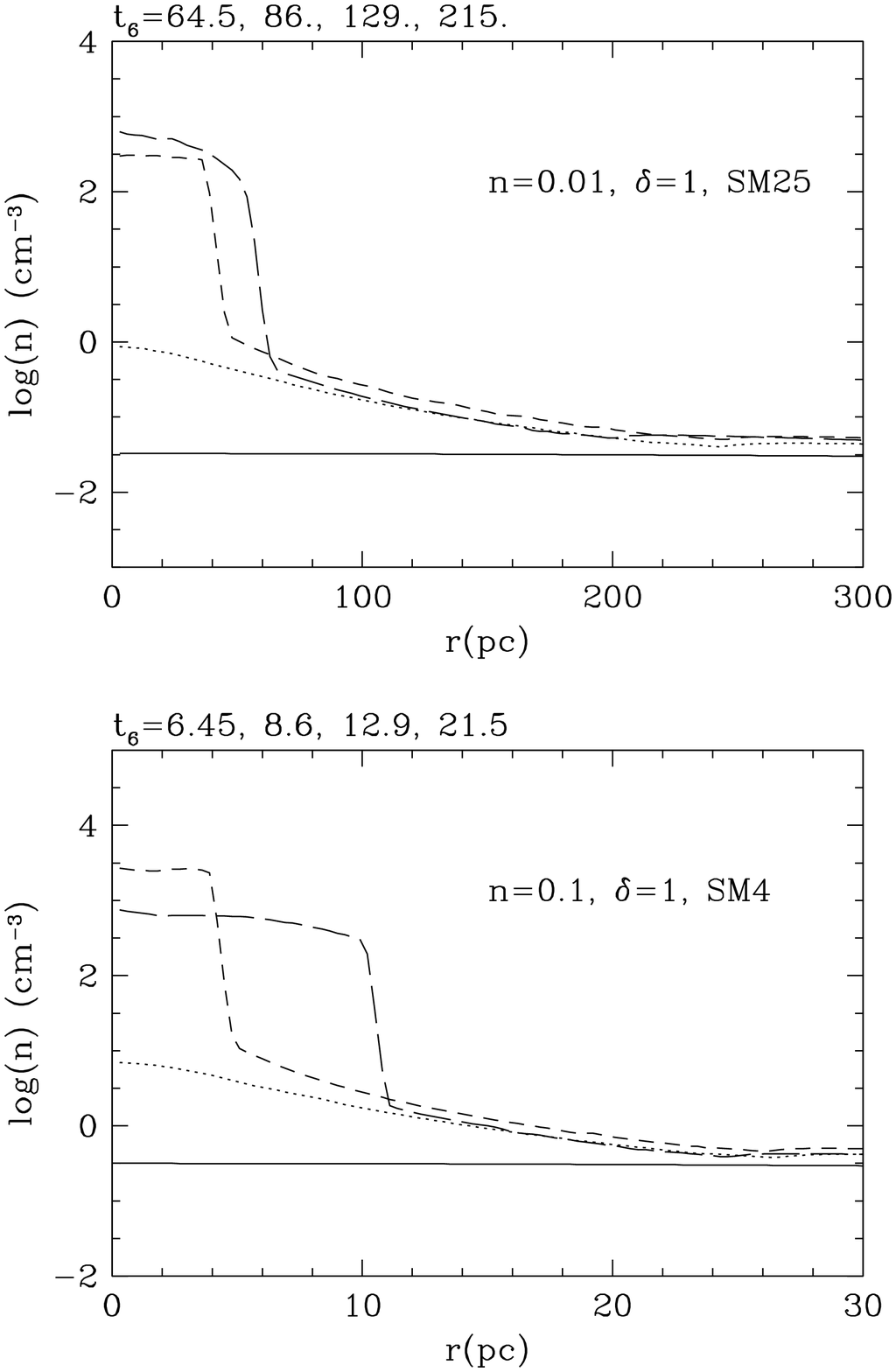}}
\vskip 0.5cm
\hskip 1.7cm {\begin{minipage}{14cm}
{\bf Fig. 4.}---~
Density profiles of PGCCs are plotted for models SM25
and SM4 at $t_6(n_h/0.1\cm3)=$ 6.45 (solid), 8.6 (dotted), 12.9
(dashed), and 21.5 (long dashed),
where $t_6$ is time in units of $10^6$ years.
They can be characterized by a core plus a $r^{-2}$ halo.
\end{minipage}}
\end{figure*}

%------------------------------------------------ FIG 4 END
Fig. 4 shows the density profiles of the
clouds of {\it intermediate} size ($R_c
\sim 0.42 l_{cool}$) with different halo densities, 
SM25 ($n_h=0.01\cm3$) and SM4 ($n_h=0.1\cm3$). 
This suggests that the density profiles of PGCCs
can be represented by a core with a halo of $\rho \propto
r^{-2}$.
Since the gravitational time scale decreases with the density
as $n^{-1/2}$ while the cooling time scale decreases as $n^{-1}$,
the relative importance of self-gravity increases in
lower density models.
For example, the model SM25 stops expanding after $t_6\sim 215$, 
while the cloud in SM4 expand outward slowly.
But the expansion speed is about $0.7 \kms$,
much less than the sound speed of $10^4 \K$ gas.

Vietri \& Pesce (1995) proposed that,
instead of the quasi-static evolution into the two-phase medium,
over-dense regions create an implosion shock
that leads to \Htwo formation and cooling
of the post-shock gas to $10^2\K$ (assuming no
external UV radiation).
However, we found that the implosion shock develops 
only if the initial density perturbation is strongly nonlinear 
($\delta >> 1$) or has a discontinuous top-hat distribution,
both of which are unlikely in real density perturbations.
Even in these cases, the implosion shock is nearly isothermal
due to the rapid cooling of the post-shock gas.
When the implosion shock rebounds at the center, it
becomes an accretion shock.
The post accretion shock gas still cools rapidly and condenses, so
Vietri \& Pesce's ``small dispersing clouds" are an artifact
of their methods.

\subsection{Physical Properties of PGCCs}

According to our numerical calculations,
only the intermediate size clouds in the supersonic compression regime,
that is, $0.4 \kpc \la R_c (n_h/0.05 \cm3) \la 2\kpc$ and
$10^{5.6} \Msun \la M_c (n_h/0.05 \cm3)^2 \la 10^8 \Msun$,
become PGCCs via thermal instability. 
Smaller clouds in the isobaric limit are not gravitationally bound
(see in next subsection), while bigger clouds in the isochoric limit
are too large to outrace the evolution of the background gas 
to become distinct structures. 
The relevant cloud sizes increase for lower halo gas densities 
(so for greater $R_g$),
since the cooling distance is inversely proportional to the gas density.
But the formation time becomes longer at lower densities and the
length scale increases to a scale on which other dynamical effects may
suppress the density growth. 

In the transition from
{\it small} to {\it intermediate} clouds, we see larger compressions owing
to the supersonic infall with the compression
factor increasing with the mass of the cloud (see Figs. 1 and 2).
Thus the final radii of PGCCs, $R_{pgcc}$, lie in the narrow range:
$R_{pgcc}~(n_h/0.05\cm3)\sim 30-60 \pc$.
This owes to the dynamics of the supersonic infall whereby
the shock in a larger cloud has a greater opportunity to accelerate
and contributes more to its overall compression.
This is extremely interesting, as the half mass radii of globular clusters
varies by only about a factor of $\sim 3$ over two decades in mass
({\it c.f.} Fall \& Rees 1977).

PGCCs form within $ 2 t_{cl,c}\approx (2\times 10^7 \yrs)$ $(0.05\cm3~/n_h)$,
while the background halo gas cools in a time scale of 
$2 t_{cl,h} \approx (8\times 10^7\yrs) (0.05\cm3/n_h)$.
So PGCCs would lose the background pressure support in the same time scale,
unless the halo gas is continuously heated by stellar winds and supernova
explosion from Pop III stars or accretion shocks or
shocks generated by merging sub-clumps. 
Thus only self-binding {\it intermediate} clouds keep their
identity, while {\small} clouds disperse.

\subsection{Fragmentation of PGCCs}

Although our numerical simulations are limited to one dimension and
cannot follow the thermal history of PGGCs below $10^4 \K$, we expect
the dense core of PGCCs may fragment further owing to self-gravity. 
If the shape of PGCCs is more like a pancake, then the
cloud would break up first by other dynamical instabilities into 
more or less spherical fragments and then only gravitationally unstable
fragments undergo further collapses.
Thus it is beyond our study to make the detail predictions on 
fragmentation processes. Here we attempt to make a rough estimate of
the minimum mass scale of unstable PGCCs.

For small isobaric clouds with $M_c < 10^{5.6} \Msun$ 
$(n_h/0.05 \cm3)^{-2}$,
the minimum unstable mass can be the ``critical mass", $M_{cr}$ adopted 
by FR85, which was derived for an isothermal sphere confined by an
external pressure (McCrea 1957),
\begin{equation} 
M_{cr}  
=1.18\left({k T_c \over {\mu m_H }}\right)^2 G^{-3/2}p_h^{-1/2}
\end{equation}
For $T_c=10^4 \K$ and $p_h=1.17\times 10^{-11} {\rm dyne~cm^{-2}}$,
$M_{cr}\sim 4.7\times 10^6 \Msun > M_{small}$, so the small clouds 
are too small to become unstable gravitationally. 

For intermediate size clouds, as shown in Figs. 3-4 (spherical cases), 
the density distribution of a typical PGCC can be approximated
by a dense core with a halo of $\rho \propto r^{-2}$. 
The flow in the core is almost static while the outer region is
infalling supersonically.
Thus the core of PGCCs can be approximated by an almost uniform static sphere.
This is certainly different from the singular isothermal distribution
($\rho \propto r^{-2}$ all the way to $r =0$), which is often assumed
in the previous studies mentioned in the introduction.
In this case, PGCCs are confined by self-gravity and
the background pressure is negligible, so $M_{cr}$ given in Eq. (14) 
is not relevant.
In a simple approach where the virial equilibrium is used (McCrea 1957)
(\eg  $2K + \Omega \approx 0$, where $2K=3MkT/(\mu m_H)$ is the thermal energy and
$\Omega = -(3/5)GM^2/R$ is the gravitational potential energy for
a uniform sphere), the minimum unstable mass is given by
\begin{equation} 
M_{vir} =5.46 \left({k T_c \over {\mu m_H G}}\right)^{3/2} 
{\rho_{c}}^{\, -1/2}, 
\end{equation}
where $T_c$ and
$\rho_c$ are the temperature and density of condensed PGCCs.
In fact this mass scale is almost the same as Jeans mass of the 
isothermal uniform sphere (Spitzer 1979) defined as
\begin{equation} 
M_J=\rho \lambda_J^3
=5.57 \left({k T_c \over {\mu m_H G}}\right)^{3/2} 
{\rho_{c}}^{\, -1/2} .
\end{equation}
Thus we take this definition of Jeans mass as the minimum mass of 
unstable PGCCs as 
$M_J \sim (5.9\times 10^6 \Msun) T_4^{3/2} n_2^{-1/2}$
where $T_4=T_c /10^4\K$ and $n_2= n_c/(100\cm3)$. 
Thus supersonic compression lowers the Jeans mass by the square
root of the additional compression factor. Owing to the spherical
convergence, the highest compression factor we have seen in our numerical
simulation is $10^6$, up to $\sim 10^{3.5}$ times higher than 
the isobaric compression. 
For PGCCs considered here, $0.1\lsim n_2\lsim 100$, so 
the Jeans mass in the {\it intermediate} clouds is 
$10^{5.5}\Msun \lsim M_J (n_h/0.05\cm3)^{1/2}$ 
$\lsim 10^{7.2}\Msun$.
Considering that the critical star formation efficiency may range
between 0.1 and 0.5 in order for the resulting cluster to be self-bounded 
(Elmegreen \etal 1999), 
this Jean mass scale can be consistent with the 
observed GC mass distribution.
If the evolution of more realistic clouds leads to
irregular pancake-like shapes (\S 2.2, BMM90), however, 
the density enhancement for real clouds would be 
smaller than what is found in the spherical simulations, 
but still larger than the isobaric compression.

The use of shocks to lower the Jeans mass
to a value that is closer to the characteristic mass
of GCs was considered by Gunn (1980) in the context
of chaotic protogalactic collapse and by Lake (1987) who invoked
shocks created by cloud-cloud collisions. The cases considered
by Gunn and Lake both have so much shear that they would be unlikely
to lead to bound structures, but shocks generic to the collapse
of the {\it intermediate} mass clouds don't have this problem 
and still serve to reduce the Jeans mass 
to a value that is characteristic of GCs.

\section{CONCLUSION}

We re-examined Fall \& Rees (1985) model for the formation of PGCCs via thermal
instabilities by numerical simulations of overdense clouds in a protogalactic
halo. The key idea of this model is that characteristic mass scale of GCs,
$M_c\sim 10^6 \Msun$, can be explained by imprinting of critical
(Jeans) mass of gas cloud at $10^4\K$ which has cooled from a hot
halo gas in pressure equilibrium via thermal instability.
If there were significant cooling below $10^4\K$ due to \Htwo and metals,
the cloud will continue to cool and the imprinting is not possible
(Kang et al 1990). Then the model fails.
So we considered the cases where the radiative cooling is
ineffective below $10^4\K$, because
the formation of \Htwo molecules is delayed due to UV radiation from a
central AGN or diffuse radiation from halo gas. 

The main conclusions are summarized as follows:
\begin{enumerate}

\item{}
Unlike X-ray cluster cooling flows, a protogalactic halo provides much
more favorable conditions for thermal instability to operate.
So PGCCs can form from marginally nonlinear (\ie $\delta \gsim 1$) density 
inhomogeneities in a protogalactic halo via thermal instability. 
Linear perturbations are likely to be disrupted by the buoyancy-driven
instabilities (Balbus \& Soker 1989; Reale \etal 1991).

\item{}
The size of thermally unstable perturbations is determined by the
cooling distance, $l_{cool}$, over which a sound wave can travel 
in a cooling time.
According to our numerical simulations in one-dimensional spherical geometry,
the clouds of $0.2 l_{cool}\la R_c \la l_{cool}$ can condense to form PGCCs
via thermal instability.
Smaller clouds ($R_c \la 0.2 l_{cool}$) cool only {\it isobarically} 
and become gravitationally unbound clouds, 
while larger cloud ($R_c \ga l_{cool}$) cool {\it isochorically} and so
do not condense to form any distinct structures.

\item{}
The thermally unstable clouds, which have mass
$10^{5.5}\Msun \la M_{PGCC}(n_h/0.05\cm3)^{2} \la10^8\Msun$,
are compressed supersonically, and the density enhancement of $10^3-10^6$
results.
This is much higher than what is expected from an isobaric compression.
Considering the larger compression factor, we estimate
the Jeans mass for PGCCs as 
$10^{5.5}\Msun \lsim M_J (n_h/0.05\cm3)^{1/2} \lsim 10^{7.2}\Msun$.
The density distribution of simulated PGGCs can be approximated by an
isothermal distribution with a constant core.

\end{enumerate}

Although we focus on the formation of old halo GCs in the Milky Way in
this work, we note globular clusters exist in some extremely different systems.
The largest of the dwarf spheroidals, Fornax, Sagittarius, NGC 147, NGC 185
and NGC 205 all possess GCs.
The specific frequency of clusters in these dwarf systems is as large as
in bright ellipticals and larger than in spirals (Hodge 1988).
If gas were in hydrostatic equilibrium in these parent galaxies,
the temperature would be just $10^4-10^5$K.
Thus a different model other than thermal instability model
for the formation of PGCCs or the formation of GCs themselves
is necessary for these systems.

Many models for GC formation have been suggested so far, but none
of them seem to be complete or can explain all the observed properties
of GCs.  One of the aspects of FR85 model that has not been touched upon
in the present work is the effect of the turbulent motions in the halo.
Turbulent energy equivalent to the virial temperature should be
injected at the largest scale and cascade down to smaller scales.
According to 2D simulations by V\'azquez-Semadeni, Gazol, \& Scalo
(2000) the turbulence cannot suppress the thermal instability,
because the time scale for the thermal instability remains shorter than the
turbulent crossing time scale ($ R_{\rm cloud}/ v_{\rm turb}$)
at smaller scales than the virialization-scale.
This is because the turbulent velocity decreases faster than the scale,
so the turbulent crossing time increases at smaller scales.
On the other hand, the turbulence can provide some heating to the halo gas.
Since the halo gas cools radiatively in about $10^8 \yrs$,
additional heating is necessary to maintain the halo temperature at $T_h$
for much longer than the cooling time scale
in order to explain the observed age spread of GCs, typically a few Gyr.

As discussed in \S2, it is necessary to consider more realistic
multi-dimensional simulations of thermally unstable clouds in a
protogalactic halo environment, which is currently under study.
The present work, based on one-dimensional considerations, however,
provides useful guidance for the up-coming multi-dimensional works.

\acknowledgments{
The work by HK was supported in part by Pusan National University 
Research Grant, 1999.
The work by GL was supported in part by NASA HPCC/ESS, ATP and LTSA 
at the University of Washington.
The work by HK and DR was supported in part by
grant 1999-2-113-001-5
from the interdisciplinary research program of the KOSEF.
HK and DR acknowledge generous hospitality by the University
of Washington, while this work had been carried out.
}


\begin{references}


\reference{} Ashman, K. M. \& Zepf, S. E. 1992, ApJ, 384, 50.

\reference{} Balbus, S. A. 1986, ApJL, 303, L79.

\reference{} Balbus, S. A. \& Soker, N. 1989, ApJ, 341, 611.

\reference{} Brinkmann, W., Massaglia, S. \& M\"uller, E. 1990, A\&A, 237, 536 (BMM90).

\reference{} Brown, J. H., Burkert, A. \& Truran, J. W. 1991, ApJ, 376, 115. 

\reference{} Colella, P. \& Woodward, P. R. 1984, J. Comp. Phys., 54, 174.

\reference{} David, L. P., Bregman, J. N. \& Seab, C. G. 1988, ApJ, 329, 66 (DBS88).

\reference{} Dopita, M. A. \& Smith, G. H. 1986, ApJ, 304, 283 .

\reference{} Elmegreen, B. G., Efremov, Y. N., Pudritz, R. E., \& Zinnecker, H., 1999,
astro-ph/9903136

\reference{} Fall, S. M. \& Rees, M. J. 1977, MNRAS, 181, 37.

\reference{} Fall, S. M. \& Rees, M. J. 1985, ApJ, 298, 18.

\reference{} Gray, D. R. \& Kilkenny, J. D., 1980, Plasma Phys., 22, 81.

\reference{} Gunn, J.E. 1980, in {\sl Globular Clusters},
ed. D. Hanes \& G. Madore, (Cambridge: Cambridge University Press), p. 301.

\reference{} Hattori, M. \& Habe, A. 1990, MNRAS, 242, 399.

\reference{} Hodge, P. 1988, PASP, 100, 568.

\reference{} Kang, H., Shapiro, P. R., Fall, S. M. \& Rees, M. J.
1990, ApJ, 363, 488.

\reference{} Kumai, Y., Baku, B. \& Fujimoto, M. 1993, ApJ, 404, 144.

\reference{} Lake, G. 1987, in {\sl IAU Symposium 127, Structure and
Dynamics of Elliptical Galaxies},
ed. T. de Zeeuw, (Dordrecht: Reidel), p. 331.

\reference{} Malagoli, A., Rosner, R. \& Bodo, G. 1987, ApJ, 319, 632.

\reference{} Malagoli, A., Rosner, R. \& Fryxell, B. 1990, MNRAS, 247, 367.

\reference{} McCrea, W. H. 1957, MNRAS, 117, 562.

\reference{} Reale, F., Rosner, R., Malagoli, A., Peres, G. \& Serio, S.
1991, MNRAS, 251, 379.

\reference{} Shapiro, P. R. \& Kang, H. 1987, ApJ, 318, 32.

\reference{} Spitzer, L. 1979, Physical Processes in the Interstellar Medium,
(New York: Wiley-Interscience).

\reference{} V\'azquez-Semadeni, E., Gazol, A., \& Scalo, J., 2000, astro-ph/0001027

\reference{} Vietri, M. \& Pesce, E. 1995, ApJ, 442, 618.

\reference{} Yoshida, T., Hattori, M. \& Habe, A. 1991, MNRAS, 248, 630.

\reference{} Zinn, R. 1985, ApJ, 392, 424.

\end{references}
\end{document}